\documentclass[aps,twocolumn,pra,showpacs,floatfix]{revtex4}
\usepackage{epsfig}
\usepackage{graphicx}
\usepackage{dcolumn}
\usepackage{amssymb,amsmath}
\usepackage{mathrsfs}
\usepackage{bm}

\begin{document}

\title{Using isotope shift for testing nuclear theory: the case of nobelium isotopes}

\author{Saleh O.  Allehabi, V. A. Dzuba, and V. V. Flambaum}

\affiliation{School of Physics, University of New South Wales, Sydney 2052, Australia}

\author{A. V. Afanasjev, and S. E. Agbemava}

\affiliation{Department of Physics and Astronomy, Mississippi State University, Mississippi 39762, USA}

\begin{abstract}
We calculate field isotope shifts for nobelium atoms using nuclear charge distributions which come from different nuclear models. We demonstrate that comparing calculated isotope shifts with experiment can serve as a testing ground for nuclear theories. It also provides a way of extracting parameters of nuclear charge distribution beyond nuclear RMS radius, e.g. parameter of quadrupole deformation $\beta$. We argue that previous interpretation of the isotope measurements in terms of $\delta \langle r^2 \rangle$ between $^{252,254}$No isotopes should be amended when nuclear deformation is taken into account. We calculate isotope shifts for other known isotopes and for hypothetically metastable  isotope $^{286}$No 
 for which the predictions of nuclear models differ substantially.

\end{abstract}


\maketitle


\section{Introduction}

Studying nuclear structure of superheavy elements (SHE) ($Z>100$) is an important area of research taking nuclear physics to unexplored territory and potentially leading to hypothetical island of stability~\cite{Rev1,Rev2,Rev3,Rev4,Rev5,GMNORSSSS.19}. The sources of experimental information are very limited since the SHE are not found in nature but produced at accelerators at very low production rate. In addition, all produced isotopes are neutron-poor and have short lifetimes (see, e.g. reviews~\cite{Rev1,Rev2,Rev3,Rev4,Rev5,GMNORSSSS.19}). Using atomic spectroscopy to measure isotope shift (IS) and hyperfine structure (hfs) is one of the promising methods to proceed. IS is widely used to extract the change of nuclear  root-mean-square
(RMS) radius between two isotopes~\cite{review_on_IS}.  In our previous paper~\cite{e120-is} we argue that it can also be used to study nuclear deformation. For example,  using different dependence of atomic transitions on nuclear structure and having IS measurements for at least two transitions, we could extract not only the change of RMS radius but also the change in quadrupole deformation parameter $\beta$. Superheavy element E120 ($Z=120$) was used in \cite{e120-is} to illustrate that if we take nuclear parameters from nuclear theory, the effect of nuclear deformation on IS is sufficiently large to be detected by modern spectroscopic methods.  The E120 element was chosen for illustration purpose because of large value of the effect. However, real measurements for E120 are not expected any time soon. The heaviest element for which IS and hfs measurements are available is nobelium ($Z=102$)~\cite{No-Nature,No-IS}. The IS is measured for $^{252,253,254}$No isotopes and hfs is measured for $^{253}$No isotope. 

 In this work we study IS of nobelium in detail. We calculate nuclear charge densities using several 
nuclear models based on covariant density functional theory \cite{VALR.05}.
Then we employ these densities in atomic calculations to get the IS and compare it to experiment. We take a closer look at the interpretation of the data and argue that nuclear deformation should be taken into account in the analysis to reduce uncertainties below 10\%.

We present a formula which expresses IS via nuclear parameters. The formula is similar to what was suggested in \cite{e120-is}. It is an analog of the standard formula IS=$F\delta \langle r^2 \rangle$ but has more terms proportional to  $\delta \langle r^2 \rangle^2$, $\Delta \beta^2$, $\Delta \beta^3$. The parameters of the formula are found from the fitting of the calculated IS. The formula is more accurate than the standard one for heavy nuclei. It can be used for predicting IS for different isotopes if nuclear parameters are taken from nuclear theory. Since the formula contains terms related to nuclear deformation, it can be used to extract the values of the change of the parameter of nuclear quadrupole deformation $\Delta \beta$ similar to how the standard formula is used to extract the change of nuclear RMS radius $\delta \langle r^2 \rangle$. IS for at least two atomic transitions is needed for this purpose. Currently IS has been measured for only one transition in nobelium. Therefore, we strongly argue in favour of new measurements and present theoretical data for three more transitions.

Finally, we make predictions for the values of the IS for some known isotopes as well as 
for the hypothetically metastable  isotope with neutron number $N=184$ which has spherical shape.

\section{Calculations}

In this work we perform nuclear and atomic calculations. Nuclear calculations are used to provide nuclear charge densities which are
connected then to observable effects, such as isotope shifts (IS) via atomic structure calculations.

\subsection{Nuclear calculations}

The nuclear properties have been calculated within the Covariant Density Functional Theory 
(CDFT) \cite{VALR.05} using several state-of-the-art covariant energy density functionals (CEDFs)
such as DD-ME2 \cite{DD-ME2}, DD-ME$\delta$ \cite{DD-MEdelta}, NL3* \cite{NL3*}, PC-PK1 
\cite{PC-PK1} and DD-PC1 \cite{DD-PC1}. In the CDFT, the nucleus is considered as a system of 
$A$ nucleons which interact via the exchange of different mesons. Above mentioned CEDFs represent 
three major classes of covariant density functional models, namely, the non-linear meson-nucleon 
coupling model (NL) [represented by the NL3* functional], the density-dependent meson exchange 
(DD-ME) model [represented by the DD-ME2 and DD-ME$\delta$ functionals] and point coupling (PC) 
model [represented by the DD-PC1 and PC-PK1 functionals]. The  main differences between them
lie in the treatment of the interaction range and density dependence. In the NL and DD-ME models, 
the interaction has a finite range which is determined by the mass of the mesons. For fixed density
it is of Yukawa type and the range is given by the inverse of the meson masses. The third class 
of models (PC model) relies on the fact that for large meson masses, the meson propagator can
be expanded in terms of this range, leading in zeroth order to $\delta$ forces and higher order 
derivative terms. Thus, in the PC model the zero-range  point-coupling interaction is used instead 
of the meson exchange \cite{VALR.05}. The NL, DD-ME and PC models typically contain 6 to 9 
parameters which  are fitted to experimental data on finite nuclei and nuclear matter properties, see 
Sec. II in Ref.\ \cite{AARR.14} for details.

Pairing correlations play an important role in all open
shell nuclei. In the present manuscript, they are taken into account
in the framework of Relativistic Hartree-Bogoliubov
(RHB) theory in which 
%
%
the RHB equations for the fermions are given by \cite{VALR.05}
\begin{eqnarray}
\begin{pmatrix}
  \hat{h}_D-\lambda  & \hat{\Delta} \\
 -\hat{\Delta}^*& -\hat{h}_D^{\,*} +\lambda
\end{pmatrix}
\begin{pmatrix}
U({\bm r}) \\ V({\bm r})
\end{pmatrix}_k
= E_k
\begin{pmatrix}
U({\bm r}) \\ V({\bm r})
\end{pmatrix}_k,
\end{eqnarray}
Here, $\hat{h}_D$ is the Dirac Hamiltonian for the nucleons with mass
$m$; $\lambda$ is the chemical potential defined by the constraints on
the average particle number for protons and neutrons;
$U_k ({\bm r})$ and $V_k ({\bm r})$ are quasiparticle Dirac
spinors \cite{VALR.05,AARR.14}
and
$E_k$ denotes the quasiparticle energies. The Dirac Hamiltonian
\begin{equation}
\label{Eq:Dirac0}
\hat{h}_D = \boldsymbol{\alpha}\boldsymbol{p} + V_0 + \beta (m+S).
\end{equation}
contains an attractive scalar potential
\begin{eqnarray}
S(\bm r)=g_\sigma\sigma(\bm r),
\label{Spot}
\end{eqnarray}
a repulsive vector potential
\begin{eqnarray}
V_0(\bm r)~=~g_\omega\omega_0(\bm r)+g_\rho\tau_3\rho_0(\bm r)+e A_0(\bm r).
\label{Vpot}
\end{eqnarray}
Since the absolute majority of nuclei are known to be axially and reflection 
symmetric in their ground states, we consider only axial and parity-conserving 
intrinsic states and solve the RHB-equations in an axially deformed harmonic 
oscillator basis~\cite{AARR.14}.  Separable pairing of finite range of Ref.\ 
\cite{TMR.09} is used in the particle-particle channel of the RHB calculations.
 
The accuracy of the description of the ground state properties (such as 
binding energies, charge radii etc) of even-even nuclei has been investigated 
globally in Refs.\ \cite{AARR.14,AA.16}. The best global description of experimental
data on charge radii has been achieved by the DD-ME2 functional [characterized by 
rms deviation of $\Delta r^{rms}_{ch}=0.0230$ fm], followed by DD-PC1 [which also provides best 
global description of binding energies], NL3* and finally by DD-ME$\delta$ 
[characterized by rms deviation of $\Delta r^{rms}_{ch}=0.0329$ fm] (see Table VI in Ref.\ \cite{AARR.14}
and Fig. 7 in Ref.\ \cite{AA.16}). However, the spread of rms deviations for charge 
radii between above  mentioned functionals is rather small  ($\Delta (\Delta r^{rms}_{ch}) = 0.0099$ fm). 
On the other hand, the charge radii of some isotopic chains (especially, those with high proton 
number $Z$) are not very accurately measured. Thus, strictly speaking we have to consider the 
accuracy of the description of charge radii by these functionals as comparable.

In the context of the study of isotopic shifts in superheavy elements  it is 
necessary to mention substantial differences in model predictions for the nuclei located 
beyond currently measured. This is contrary to the fact that nuclear theories in general 
agree  on the properties  of SHE which have already been measured in experiment 
(see,  for example, Figs. 7 and 8 in Ref.\ \cite{Afanasiev}).  For example, some CEDFs 
(such as NL3*, DD-ME2 and PC-PK1) predict a band of 
spherical nuclei along and near the proton number $Z=120$ and neutron number 
$N=184$ [see Figs. 6 (a), (b) and (e) in Ref.\ \cite{Afanasiev}].  However, for other
functionals (DD-PC1 and DD-ME$\delta$) oblate deformed shapes dominate at and 
in the vicinity of these lines [see Figs. 6 (c) and (d) in Ref.\ \cite{Afanasiev}]. Nuclear
measurements of the energies of the excited $2^+$ states are needed to discriminate 
experimentally between spherical and oblate deformed ground states. Such 
experiments are not possible nowadays. It would be interesting to see whether
atomic measurements  would be able to help with such a discrimination.

\subsection{Atomic calculations}

Nuclear calculations produce nuclear charge density as a two-dimensional function
$\rho(z,r_{\perp})$, where $z$ is the coordinate along the axis of symmetry and
$r_{\perp}$ is the radial coordinate in the direction perpendicular to the axis of symmetry.
Atomic electrons feel the nucleus as a spherically-symmetric system, averaged over nuclear rotations. 
Therefore, we transform $\rho(z,r_{\perp})$ into spherical coordinates $\rho(r,\theta)$ and average it over $\theta$, 
$\rho(r) = \int \rho(r,\theta)d\theta$. The density $\rho$ is normalized by the
condition $\int \rho dV = Z$. In the end we have nuclear charge density in numerical form rather than a set of parameters as in the case of using standard Fermi distribution. However, it is often useful to have such parameters as nuclear root-mean square (RMS) radius $R_p$, parameter of quadrupole deformation $\beta$, etc. Having them allows to study the sensitivity of observable effects (isotope shift in our case) to the change in the values of these parameters. It turns out that the IS is most sensitive to the change of $R_p$ and $\beta$. We restrict our discussion to these two parameters. We find their values by integrating nuclear charge density. 

IS can be found by direct comparison of the calculations for two different isotopes. This works well for isotopes which differ by large number of neutrons, $\Delta N \gg 1$. For neighbouring isotopes, where $\Delta N \sim 1$, the IS is small and its calculated value is affected by numerical uncertainties.
 To suppress numerical noise we use so-called finite field method~\cite{FFM}. We construct nuclear potential according to the formula $V_N = V_1 +\lambda (V_2 - V_1)$, where $V_1$ and $V_2$ are nuclear potentials for two isotopes and $\lambda$ is numerical factor which can be large to enhance the difference between two isotopes and thus suppress the numerical noise. First, the calculations are done for $\lambda=0$ to obtain reference transition frequencies. Then, they
are performed for several values of $\lambda >1$ and the frequencies are extrapolated to $\lambda=1$. In practice, we use $\lambda=2$ and $\lambda=4$.

To perform electron structure calculations
we start from the so-called CIPT method (Configuration Interaction with Perturbation Theory)~\cite{cipt}. It treats No as a system with 16 external electrons allowing excitations from the $5f$ subshell into the CI space. The aim of this study is to check whether the mixing of the $4f^{14}7snp$ ($n=7,8$) and $4f^{13}7^2s6d$ configurations has any significance for our states of interest.  Such study was performed before~\cite{No-IS,No-hfs} for the lowest odd states of No, $7s7p \ ^3$P$^{\rm o}_1$ and  $^1$P$^{\rm o}_1$. The answer was negative. Now we want to extend our study to two more states $7s8p \ ^3$P$^{\rm o}_1$ and  $^1$P$^{\rm o}_1$. Therefore, we performed the CIPT calculations again and found that there is no strong mixing of our states of interest with the state involving excitations from the $5f$ shell. This means that No can be treated as an atom with two valence electrons above closed shells. We use well established CI+MBPT method~\cite{CI+MBPT,Ba-CI} to perform the calculations.

The effective CI hamiltonian has a form
\begin{equation}
H^{\rm CI}(r_1,r_2) = \hat h_1(r_1) + \hat h_1(r_2) + \frac{e^2}{r_{12}} + \Sigma_2(r_1,r_2),
 \label{eq:HCI}
\end{equation}
where $\hat h_1$ is the single-electron part of the Hamiltonian, which is the sum of the Hartree-Fock operator $\hat H^{\rm HF}$ and correlation potential $\Sigma_1$, $\hat h_1 = \hat H^{\rm HF} + \Sigma_1$. Correlation potential $\Sigma_1$ is an operator which includes correlations between a particular valence electron and the electrons in the core. The operator $\Sigma_2$ can be understood as screening of Coulomb interaction between valence electrons by core electrons. We calculate $\Sigma_1$ and $\Sigma_2$ in second order of the many-body perturbation theory. The contribution of higher orders is relatively small but not totally negligible~\cite{No-hfs,SDS+CI,SDme+CI}. To simulate them, we rescale the $\Sigma_1$ operator in the $s$ and $p$-waves to fit the known energy of the $^1$S$_0 - ^1$P$^{\rm o}_1$ transition,
$\Sigma_1(s) \rightarrow 0.8 \Sigma_1(s)$, $\Sigma_1(p) \rightarrow 0.94 \Sigma_1(p)$. The rescaling helps to make more accurate 
predictions for the positions of other odd levels. It also improves the wave functions used to calculate transition amplitudes.

We perform the calculations of the electric dipole transition rates between the ground and four lowest in energy odd states
 to see whether the rates are sufficiently large for the measurements. We use random-phase approximation (RPA) for the calculations. The RPA  equations for the core states have a form
\begin{equation}\label{e:RPA}
(\hat H^{\rm HF} + \epsilon_c)\delta \psi_c =-(d+\delta V)\psi_c,
\end{equation}
where $d$ is electric dipole operator, the index $c$ numerates the states in the core, $\delta \psi_c$ is 
the correction to the core orbital caused by external electric field and $\delta V$ is the correction to the self-consistent HF potential caused by the change of all core states. The RPA equations are solved self-consistently for all states in the core. As a result, we have $\delta V$  which is used to calculate transition amplitudes between valence states
\begin{equation} \label{e:E1}
 A_{ab} = \langle a | d + \delta V| b \rangle.
\end{equation}
Here $a$ and $b$ are two-electron wavefunctions found in the CI+MBPT calculations. The rate of spontaneous decay of the state $b$ into the state $a$ via an electric dipole transition is given by (in atomic units)
\begin{equation}
T_{ab} = \frac{4}{3}(\omega_{ab} \alpha)^3\frac{A_{ab}^2}{2J_b+1}.
\end{equation}

\section{Results}

\subsection{Energies and transition rates}

\begin{table}
\caption{\label{t:energies} Excitation energies, electric dipole transition amplitudes and rates of spontaneous decay via electric dipole transitions to the ground state for four odd states of nobelium.}
\begin{ruledtabular}
\begin{tabular}{l rrrdr}
\multicolumn{1}{c}{Upper}&
\multicolumn{3}{c}{Excitation energies [cm$^{-1}$]}&
\multicolumn{1}{c}{$A_{ab}$}& 
\multicolumn{1}{c}{Transition}\\
\multicolumn{1}{c}{state}&
\multicolumn{1}{c}{Present}&
\multicolumn{1}{c}{Expt.~\cite{No-Nature}}&
\multicolumn{1}{c}{CI+all~\cite{No-hfs}}& 
\multicolumn{1}{c}{[a.u.]}& 
\multicolumn{1}{c}{rate [s$^{-1}$]}\\ 
\hline
$7s7p \ ^3$P$^{\rm o}_1$ & 21213 &            & 21042 & 1.37 & $1.2\times 10^7$ \\
$7s7p \ ^1$P$^{\rm o}_1$ & 29963 & 29961 & 30203 &  4.24 & $3.3\times 10^8$  \\
$7s8p \ ^3$P$^{\rm o}_1$ & 41482 &            &            &  0.097 & $3.6\times 10^5$ \\
$7s8p \ ^1$P$^{\rm o}_1$ & 42926 &            &            &  0.86 & $4.0\times 10^7$ \\
\end{tabular}
\end{ruledtabular}
\end{table}
The results of calculations for the energies and transition rates are presented in Table~\ref{t:energies}. Good agreement with experiment is the result of fitting. The {\em ab initio} CI+MBPT result for the energy of the $7s7p \ ^1$P$^{\rm o}_1$ state is 31652~cm$^{-1}$. This value differs from experimental one by 5.6\%. Comparing it with the CI+all-order result of Ref.~\cite{No-hfs} shows that  most of this difference is due to higher-order correlations.

The $7s7p \ ^1$P$^{\rm o}_1$ state has the largest electric dipole transition amplitude and largest transition rate to the ground state. There are at least two more transitions (first and last lines of Table~\ref{t:energies}) which are probably strong enough to be experimentally studied. Note, that at least two transitions are needed to use isotope shift to extract nuclear deformation (see below). 

\subsection{Comparing nuclear models}

\begin{figure}[tb]
\epsfig{figure=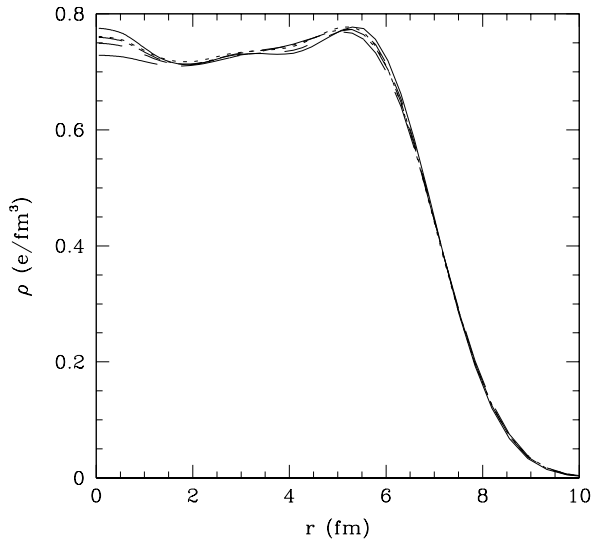,scale=0.4}
\caption{Symmetrized nuclear densities in five nuclear models considered in this work. See Fig.~\ref{f:models1} for details.}
\label{f:models}
\end{figure}
\begin{figure}[tb]
\epsfig{figure=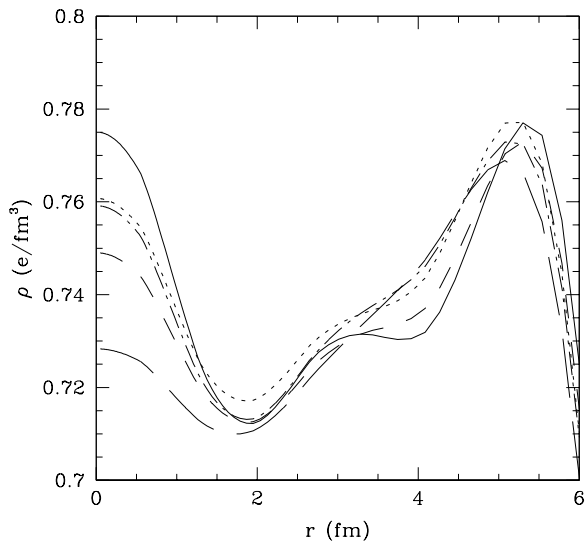,scale=0.4}
\caption{Upper left part of  Fig.~\ref{f:models} showing the details of nuclear density in five nuclear models. Solid line - DD-ME$\delta$, dot line -  DD-ME2, short dash line - NL3*, long dash line - PC-PK1, dot-short dash line - DD-PC1.}
\label{f:models1}
\end{figure}

\begin{table*}
\caption{\label{t:models}Parameters of nuclear model and corresponding calculated isotope shift for the $7s^2 \ ^1$S$_0$ - $7s7p \ ^1$P$^{\rm o}_1$ transition in $^{252,254}$No. $R_p$ is nuclear RMS charge radius ($R_p = \sqrt{\langle r^2 \rangle}$), $\beta$ is a parameter of nuclear quadrupole deformation, IS is calculated isotope shift, $F$ is field shift constant ($F={\rm IS}/\delta \langle r^2\rangle$). Here $\Delta \beta =\beta_1-\beta_2$; the subscripts 1/2 correspond  to the isotope with higher/lower
value of neutron number. } 
\begin{ruledtabular}
\begin{tabular}{ldddd dddd}
\multicolumn{1}{c}{Nuclear}&
\multicolumn{2}{c}{$^{252}$No}&
\multicolumn{2}{c}{$^{254}$No}&
\multicolumn{1}{c}{$\delta \langle r^2\rangle$}&
\multicolumn{1}{c}{$\Delta \beta$}&
\multicolumn{1}{c}{IS}&
\multicolumn{1}{c}{$F$}\\
\multicolumn{1}{c}{Model}&
\multicolumn{1}{c}{$R_p$ [fm]}&
\multicolumn{1}{c}{$\beta$ }&
\multicolumn{1}{c}{$R_p$ [fm]}&
\multicolumn{1}{c}{$\beta$ }&
\multicolumn{1}{c}{fm$^2$}&&
\multicolumn{1}{c}{cm$^{-1}$}&
\multicolumn{1}{c}{cm$^{-1}/{\rm fm}^2$}\\
\hline
 DD-ME2  & 5.97171 & 0.298 & 5.98349 & 0.298 & 0.1408 & 0.000 & -0.482 & -3.42 \\
 DD-ME$\delta$  & 5.96390 & 0.284 & 5.97259 & 0.278 & 0.1037 & 0.006 & -0.374 & -3.61 \\
 NL3s      & 5.97447 & 0.300 & 5.98772 & 0.298 & 0.1585 & 0.002 & -0.503 & -3.17 \\
 PC-PK1   & 5.98639 & 0.306 & 5.99967 & 0.305 & 0.1592 & 0.001 & -0.538 & -3.38 \\
 DD-PC1   & 5.97208 & 0.297 & 5.98225 & 0.295 & 0.1216 & 0.001 & -0.431 & -3.54 \\
\end{tabular}
\end{ruledtabular}
\end{table*}

Figs.~\ref{f:models} and \ref{f:models1} show symmetrised nuclear densities ($\rho(r) = \int \rho(r,\theta)d\theta$) for nuclear models used in this work.
Table~\ref{t:models} shows the parameters of nuclear charge distribution for these models (CEDFs) and corresponding calculated isotope shifts for the $7s^2 \ ^1$S$_0$ - $7s7p \ ^1$P$^{\rm o}_1$ line of  $^{252}$No and $^{254}$No. Experimental value for the isotope shift is 0.336(23)~cm$^{-1}$~\cite{No-IS}. The DD-ME$\delta$ model leads to the best agreement of the calculated and experimental IS; the calculated value is only about 10\% larger then the experimental one. Note also that this model predicts the largest value of $\Delta \beta$ between two isotopes. Last column of Table~\ref{t:models} presents the ratios of calculated isotope shift to $\delta \langle r^2\rangle$, which is the field shift constant $F$. In the absence of nuclear deformation this constant should not depend on nuclear model, i.e. it should be the same everywhere. However, we see that it varies significantly. This is an indication that nuclear deformation is important. In Ref.~\cite{No-IS} nuclear field constant $F$ was calculated without taking into account nuclear deformations. The CI+MBPT value of Ref.~\cite{No-IS} is -3.47~cm$^{-1}$/fm$^2$. It is in excellent agreement with our value -3.42~cm$^{-1}$/fm$^2$ obtained with the same method and with the use of the DD-ME2 nuclear model in which $\Delta \beta=0$ for the two isotopes (see Table~\ref{t:models}). However, the calculations of IS based on this model overestimate IS by about 40\%. 
If we assume that the overestimation of the IS mostly comes from the overestimation of 
$\delta \langle r^2 \rangle$, then the DD-ME$\delta$ results provide more consistent picture.
Indeed, the transition from the DD-ME2 to DD-ME$\delta$ model leads to  the reduction of 
$\delta \langle r^2 \rangle$ from 0.1408~fm$^2$ down to 0.1037~fm$^2$ (see Table~\ref{t:models}). The latter 
value is very close to  $\delta \langle r^2 \rangle=$0.105(7)(7)~fm$^2$ found in Ref.~\cite{No-IS}.
In addition, the calculated IS of the DD-ME$\delta$ model of $-0.374$~cm$^{-1}$ (see Table~\ref{t:models})
is very close to the experimental value of $-0.336(23)$~cm$^{-1}$ (see Ref.\ \cite{No-IS}).
Note that the best agreement with experiment is achieved with the nuclear model which gives the largest change in nuclear deformation
parameter between two isotopes. This indicates the importance of taking nuclear deformation into account in the analysis.

\subsection{Using isotope shift measurements to find parameters of nuclear charge distribution}

\begin{table*}
\caption{\label{t:abcde} The parameters of formula (\ref{e:abcde}) for isotope shifts from the ground state ($7s^2 \ ^1$S$_0$) to excited odd states of nobelium.}
\begin{ruledtabular}
\begin{tabular}{lddddd}
\multicolumn{1}{c}{Odd states}&
\multicolumn{1}{c}{$F$}&
\multicolumn{1}{c}{$G$}&
\multicolumn{1}{c}{$a$}&
\multicolumn{1}{c}{$b$}&
\multicolumn{1}{c}{$c$}\\
&\multicolumn{1}{c}{cm$^{-1}$/fm$^2$}&
\multicolumn{1}{c}{cm$^{-2}$/fm$^4$}&
\multicolumn{1}{c}{cm$^{-1}$}&
\multicolumn{1}{c}{cm$^{-1}$}&
\multicolumn{1}{c}{cm$^{-1}$/fm$^2$}\\
\hline
$7s7p \ ^3$P$^{\rm o}_1$ &  -3.7828   &   0.0288  &   -1.4013  &    1.3708  &   -0.0215 \\
$7s7p \ ^1$P$^{\rm o}_1$ &  -3.5042   &   0.0254   &  -1.2247  &    1.2234  &   -0.0152 \\
$7s8p \ ^3$P$^{\rm o}_1$ &  -3.2063   &   0.0265  &   -1.0941  &    1.1304  &   -0.0071 \\
$7s8p \ ^1$P$^{\rm o}_1$ &  -3.3112   &   0.0245   &  -1.1592   &   1.1919   &  -0.0090 \\
\end{tabular}
\end{ruledtabular}
\end{table*}

\begin{table*}
\caption{\label{t:IS184}Isotope shifts between $^{254}$No and $^{286}$No in different nuclear models for four electric dipole transitions from the ground state (cm$^{-1}$).}
\begin{ruledtabular}
\begin{tabular}{l dddddd}
\multicolumn{1}{c}{Nuclear}&
\multicolumn{1}{c}{$R_p$ for $^{286}$No}&
\multicolumn{1}{c}{$\delta \langle r^2 \rangle$}&
\multicolumn{4}{c}{Upper state}\\
\multicolumn{1}{c}{model}&
\multicolumn{1}{c}{[fm]}&
\multicolumn{1}{c}{[fm$^2$]}&
\multicolumn{1}{c}{$7s7p \ ^3$P$^{\rm o}_1$ } &
\multicolumn{1}{c}{$7s7p \ ^1$P$^{\rm o}_1$ } &
\multicolumn{1}{c}{$7s8p \ ^3$P$^{\rm o}_1$ } &
\multicolumn{1}{c}{$7s8p \ ^1$P$^{\rm o}_1$ } \\
\hline
DD-ME2              & 6.084420 & 1.1872 & -4.52       & -4.18 & -3.84 & -3.97 \\
DD-ME$\delta$   & 6.075497 &  1.2111 & -4.61       & -4.27 & -3.90 & -4.03 \\
NL3*                    & 6.097316 &  1.3029 & -4.94       & -4.57 & -4.20 & -4.34 \\
PC-PK1        & 6.114652 &  1.3655 & -5.17       & -4.78 & -4.39 & -4.54 \\
DD-PC1               & 6.085116 &  1.2212 & -4.64       & -4.29 & -3.95 & -4.08 \\
Average   &                 &   1.2576   & -4.78(40) & -4.42(36) & -4.06(33) & -4.19(35) \\
\end{tabular}
\end{ruledtabular}
\end{table*}

It was suggested in our previous work~\cite{e120-is} to fit field isotope shift between two isotopes with the formula which depends on the change of two nuclear parameters, nuclear RMS radius, and quadrupole deformation parameter $\beta$. 
Here we present the formula in slightly different form 
\begin{eqnarray}\label{e:abcde}
 &&\delta \nu = F\delta\langle r^2\rangle +G (\delta \langle r^2\rangle)^2 + \\
&& a \Delta(\beta^2) +b \Delta(\beta^3) + c \delta\langle r^2\rangle \Delta(\beta^2) \nonumber 
\end{eqnarray}
Here $\delta\langle r^2\rangle = \langle r^2\rangle_1 - \langle r^2\rangle_2$ is  the change of square of nuclear RMS radius, $\Delta(\beta^2) = \beta_1^2-\beta_2^2$, $\Delta(\beta^3) = \beta_1^3-\beta_2^3$
 and the indexes $1$ and $2$ numerate isotopes, index $1$ corresponds to an isotope with higher value of the neutron number.
The coefficients $F, G, a, b,c$ in this formula are found by least squares fitting of calculated IS for a wide range of nuclear parameters.
The values of these parameters for four electric dipole transitions in nobelium are presented in Table~\ref{t:abcde}. Note that the value of $F$ for the second transition is in excellent agreement with the CI+MBPT calculations of Ref.~\cite{No-IS}.

First term in Eq. (\ref{e:abcde}) represents standard formula for field IS. It ignores nuclear deformation and relativistic corrections.
It was shown in Ref.~\cite{F-IS} that relativistic effects make the field constant $F$ 
isotope-dependent. It was suggested to use a modified formula $\delta \nu_i = F^{\prime} \delta \langle r^{2\gamma} \rangle$, where $\gamma = \sqrt{1-(z\alpha)^2}$. Modified field shift constant $F^{\prime}$ does not depend on isotopes. However, this formula works well only for spherical nuclei~\cite{e120-is}. In contrast, formula (\ref{e:abcde}) can be used for a wide range of nuclei. Relativistic corrections in it are fitted with  quadratic in $\delta\langle r^2\rangle $ term (second term in (\ref{e:abcde})).  This formula can be used to predict IS for different isotopes and atomic transitions if nuclear parameters are taken from nuclear theory. 

The formula can also be used in an opposite way: the change of nuclear parameters can be found from IS measurements. Since formula (\ref{e:abcde}) depends on two nuclear parameters, the measurements of IS for at least two atomic transitions are needed. Then standard mathematical procedures can be used to solve the system of two quadratic equations to find the change of nuclear parameters.

 For neighbouring isotopes second and last terms in Eq. (\ref{e:abcde}) can be neglected (see Table \ref{t:abcde})
and remaining terms reduced to
\begin{equation}\label{e:Fb}
\delta \nu = F\delta\langle r^2\rangle +d \Delta \beta.
\end{equation}
The parameters $F$ and $d$ in this formula are isotope-dependent and should be calculated for one of the considered isotope.
The parameter $d$ is related to $a$ and $b$ in (\ref{e:abcde}) by $d = a(\beta_1+\beta_2)+b(\beta_1^2+\beta_1\beta_2+\beta_2^2)$
and $\Delta \beta =\beta_1-\beta_2$.

So far the IS has been measured for one transition (second transition in Table~\ref{t:abcde}) between isotopes $^{252,253,254}$No. According to nuclear theory~\cite{Afanasiev}, all these isotopes have deformed 
shapes, e.g. for $^{252,254}\Delta \beta =0.006$  for DD-ME$\delta$ CEDF
(see Table~\ref{t:models}). Using the formula (\ref{e:Fb}) and the numbers from Table~\ref{t:abcde} we find that the contribution of the second term in (\ref{e:Fb}) into IS is 0.003~cm$^{-1}$. This is 8 times smaller than the uncertainty of the measurements (measured value for IS is 0.336(23)~cm$^{-1}$~\cite{No-IS}). Therefore, to see the effect of nuclear deformation one has to either increase the accuracy of the measurements or use different isotopes. Note also that the measurements need to be done for at least two atomic transitions. Currently, IS is measured only for one transition in No~\cite{No-IS}.

   Finally, we calculated isotope shifts between the $^{254}$No and $^{286}$No isotopes in different nuclear models; 
the results are presented in Table~ \ref{t:IS184}.  Note that  the $^{286}$No nucleus has neutron number $N=184$ which is 
magic number  in this mass region \cite{Afanasiev,GMNORSSSS.19}  corresponding to a large shell closure. Thus, according to
nuclear theory this nucleus has spherical shape. It is expected to be long-living isotope \cite{GMNORSSSS.19}. 
One transition frequency has been already measured in the $^{254}$No isotope~\cite{No-Nature}. 
One can use the isotope shift from Table~\ref{t:IS184} to correct measured frequencies of atomic transitions from  $^{254}$No to $^{286}$No isotopes and use the data for 
a search of long-living No isotopes in astrophysical data~\cite{astro}. Note that all nuclear models give very close predictions for the IS (see Table~\ref{t:IS184}). We 
use the spread of calculated results for an estimation of the uncertainties in the predictions and an average calculated value as the central point of these predictions.

\subsection{Nuclear deformation and nonlinearity of King plot}

It was suggested in Ref.~\cite{non-l-King} to use possible nonlinearity of King plot to search for new particles. If some presently unknown bosons mediate interaction between atomic electrons and neutrons in the nucleus, then field shift constant $F$ would depend on the number of neutrons. This would manifest itself in the nonlinearity of King plot. Let us consider how this consideration is affected by nuclear deformation. The only condition for the King plot to be linear is the separation of nuclear and electron variables. 
Let us consider standard formula for isotope shift, namely, $\delta \nu = F\delta\langle r^2 \rangle + M N$. Here $F$ is field shift constant, $\delta\langle r^2 \rangle$ is a nuclear factor describing change in nuclear structure between two isotopes, $M$ is mass factor $M=(M_b-M_a)/M_bM_a$, $N$ is electron structure factor related to mass shift, and the indexes $a$ and $b$ numerate isotopes. 
 If $F$ does not depend on nucleus and $\delta\langle r^2 \rangle$ does not depend on electrons then one can write for two atomic transitions
\begin{equation}
(\delta \nu_1/M) = \frac{F_1}{F_2}(\delta \nu_2/M) +\frac{F_1}{F_2}N_2 + N_1.
\label{e:King}
\end{equation}
One can see that on the $\delta \nu_1/M, \delta \nu_2/M$ plane the points corresponding to different isotopes are all on the same line.
If formula (\ref{e:Fb}) is used for the field shift then extra term appears in (\ref{e:King})
\begin{eqnarray}
(\delta \nu_1/M) &=& \frac{F_1}{F_2}(\delta \nu_2/M) +\frac{F_1}{F_2}N_2 + N_1 +\\
&+& \frac{\Delta \beta}{M}\left(d_1-\frac{F_1}{F_2}d_2\right). \nonumber
\label{e:Kingm}
\end{eqnarray}
This last term does depend on isotopes and thus breaks the linearity of King plot. It is instructive to see when this term is zero. The most obvious case is $\Delta \beta=0$, i.e. all considered isotopes have the same nuclear deformation. This is unlikely scenario for heavy nuclei. However, the terms can be small if deformations are similar.
 The less obvious case is $d_1-d_2F_1/F_2=0$. Note that the expression $d_1F_2-d_2F_1$ is the determinant of the system of two linear equations for $\delta \langle r^2 \rangle$ and $\Delta \beta$ if IS for two transitions is given by Eq.
  (\ref{e:Fb}). The determinant is zero means that the equations are proportional to each other and cannot be resolved. This might be the case of the transitions between similar states, e.g. $7s - 7p_{3/2}$ and $7s - 8p_{3/2}$ transitions in No$^{+}$. Exact proportionality is unlikely but strong suppression is possible (i.e $d_1F_2 \approx d_2F_1$). The suppression is less likely in many-electron atoms since  the states are affected by configuration mixing and it is different for low and high energy states so that similar transitions can hardly be found. 
 
\acknowledgments

This work was funded in part by the Australian Research Council. The material is based upon 
work supported by the U.S. Department of Energy, Office of Science, Office of Nuclear Physics 
under Award No. DE-SC0013037.
 

\end{document}